\documentclass[%
twocolumn,
groupedaddress,
amsmath,amssymb, nofootinbib
]{revtex4-2}

\usepackage[english]{babel}

\usepackage{algorithm2e}
\usepackage{natbib}
\usepackage{color}

\usepackage{amsmath}
\usepackage{amssymb}
\usepackage{graphicx}
\usepackage{caption}
\usepackage[mathscr]{euscript}
\usepackage[colorlinks=true, allcolors=blue]{hyperref}

\begin{document}

\title{Fluid dynamics meet network science: two cases of temporal network eigendecomposition}
\author{Lucas Lacasa}
\email{lucas@ifisc.uib-csic.es}
\affiliation{Institute for Cross-Disciplinary Physics and Complex Systems (IFISC, CSIC-UIB),\\Campus UIB, 07122 Palma de Mallorca, Spain}


\begin{abstract}
Temporal networks, defined as sequences of time-aggregated adjacency matrices, sample latent graph dynamics and trace trajectories in graph space. By interpreting each adjacency matrix as a different time snapshot of a scalar field, we show how fluid-mechanics methods can be applied to construct two distinct eigendecompositions of temporal networks. The first builds on the proper orthogonal decomposition (POD) of flowfields and decomposes the evolution of a network in terms of a basis of orthogonal network eigenmodes which are ordered in terms of their relative importance, hence enabling compression of temporal networks as well as their projection in low-dimensional embeddings. The second proposes a numerical approximation of the Koopman operator, a linear operator acting on a suitable observable of the graph space which provides the best linear approximation of the latent graph dynamics. Its eigendecomposition provides a data-driven spectral description of the temporal network dynamical stability, in terms of dynamic modes which grow, decay or oscillate over time. We illustrate and validate the application of both eigendecompositions in a suite of synthetic generative models of temporal networks with varying complexity.
\end{abstract}

\maketitle

\section*{Introduction}
Network science \cite{latora2017complex} models how individual elements interact with each other in a complex system, and how such architecture backbone conditions and influences the emergence of complex dynamical phenomena. When interactions also fluctuate over time, networks (or graphs) are themselves time-varying, leading to the theory of temporal networks \cite{holme2012temporal, masuda2016guide}. Among other possible definitions, in this work we call a Temporal Network (TN) a (time) ordered sequence of graphs. While complexity science has predominantly investigated how dynamics defined on top of network structures are affected when such structures are time-varying \cite{zanin2009dynamics, van2013non, masuda2013temporal,scholtes2014causality,delvenne2015diffusion,williams2019effects}, relatively lesser work has considered principled analysis of the intrinsic dynamics of temporal networks \cite{williams2022shape}. In this regard, a recent approach considers TNs as time series of graphs --i.e. trajectories in graph space of a possibly latent graph dynamics--, paving the way for extensions to the network realm of classical dynamical systems and time series analysis concepts such as temporal autocorrelation \cite{lacasa2022correlations, bauza2023characterization, andres2024detecting, hartle2025autocorrelation}
or Lyapunov exponents \cite{annalisa, danovski2024dynamical, jimenez2025leveraging, caligiuri2026predictability}.\\
Here we take this latter view and further interpret the sequence of adjacency matrices which defines a TN as sampling a discrete scalar field whose evolution is given by a latent graph dynamics \cite{prisner1995graph}.
This, we shall see, naturally enables the use of tools from fluid mechanics \cite{taira2017modal} to characterise such network trajectories and to approximate the latent graph dynamics. {We illustrate this proposal by building on two different fluid-mechanic methodologies, namely proper orthogonal decomposition (POD) \cite{lumley1967structure} and dynamic mode decomposition (DMD) \cite{schmid2010dynamic}, the latter being one possible numerical implementation of the Koopman operator \cite{koopman1931hamiltonian, mezic2013analysis, mezic2005spectral}. We show how to interpret these techniques in terms of two mathematically-principled eigendecompositions of temporal networks, and illustrate how these could be leveraged for temporal network reconstruction by focusing on two different aspects:} POD exploits finding low-rank embeddings \cite{thongprayoon2023embedding, lacasa2025scalar} to {\it compress} the TN's structure, whereas DMD focuses on the network's dynamical {\it stability} --and its subsequent control \cite{liu2011controllability, li2017fundamental}--.\\
{Before presenting the theory and results of its implementation to a range of different TNs, it is important to disclaim that both the compression and the stability aspects of temporal networks are, in their own right, important problems that have been addressed previously in the network science and computer science literature. For instance, a variety of different strategies have been proposed over recent years to compress the information stored in a temporal network, using both linear and nonlinear dimensionality reduction techniques \cite{thongprayoon2023embedding, lacasa2025scalar, Nguyen2018ContinuousTimeDN, LiNetworkEmbedding} and unsupervised clustering methods \cite{masuda2019detecting, li2025revisiting}. Likewise, the linear stability of the network evolution and its associated prediction problem has been explored recently, using a range of methods encompassing state-space models and machine learning, to cite some \cite{shvydun2023system, he2024sequential, li2024dyg}.
In this sense, we shall clarify that our goal in this work is not necessarily to benchmark brand new data-driven methods against SOTA on compression and stability tasks. Our aim here is rather to conceptualise and provide proofs of concept on a potentially fruitful and underexplored connection between fluid dynamics and network science. As a matter of fact, whereas some authors have investigated how network-theoretic tools can be exploited to extract meaningful insights in fluid dynamic problems \cite{taira2016network, ser2015flow, ser2021lagrangian, iacobello2019lagrangian, iacobello2021review, iacobello2021large}, the opposite direction has seldom be explored. Here we present evidence that such connection is worth pursuing. To illustrate such connection, we focus on the two concrete problems of network reconstruction via network compression and network stability and show how application of two well-established fluid-dynamic techniques can offer interesting insights. Our hope is that this paper stimulates further research deepening the cross-pollination between fluid dynamics and network science.}

\section*{THEORY}
A temporal network (TN) is defined here as an ordered sequence of $m$ real-valued adjacency matrices $({\bf A}(1),{\bf A}(2),\dots,{\bf A}(m))$, where ${\bf A}({t})\in \mathbb{R}^{n \times n}$ aggregates the activity of a latent graph dynamics between $t-1$ and $t$. {This is often called the graph (or network) {\it snapshot} representation of a TN, where each of the adjacency matrices in the TN sequence --each graph snapshot-- builds a static aggregation of the TN over a short period of time}.
Let us suitably flatten each adjacency matrix into a vector 
\begin{equation}
{\bf a}(t)=\texttt{flat}({\bf A}({t}) - \langle {\bf A} \rangle),
\label{eq:flatten}
\end{equation}
 where $\langle {\bf A} \rangle= \frac{1}{m}\sum_{t=1}^m {\bf A}(t)$ is the time-averaged adjacency matrix (sometimes called the annealed adjacency matrix), and flattening is such that $\texttt{flat}({\bf A})=(A_{11},A_{12},\dots,A_{1n},A_{21},A_{22},\dots, A_{2n},\dots,A_{n1},\dots,A_{nn})$. Accordingly, ${\bf a}(t)\in \mathbb{R}^{n^2}$ is a real-valued column vector\footnote{For simple, undirected networks, $n^2$ reduces to $n(n-1)/2$, and for unweighted graphs, entries are binary.}. By subsequently stacking each of the $m$ flattened vectors in order, the TN  admits a $n^2 \times m$ matrix representation ${\mathscr{S}}$ such that
\begin{equation}
{\mathscr{S}}\equiv {\mathscr{S}}_{[1\dots m]}:=[{\bf a}(1) \ {\bf a}(2) \ \dots \ {\bf a}(m)]. 
\label{eq:matrix}
\end{equation}
In this way, each network snapshot is nothing but a point in a $n^2$-dimensional Euclidean space.\\
Our first goal is to be able to project each (normalised) network snapshot into a space spanned by a base of only $r$ (ideally, with $r\ll n^2$) orthogonal network modes $\{\phi_j\}_{j=1}^r$ such that
\begin{equation}
{\bf a}(t) \approx \sum_{j=1}^r \alpha_j(t) \phi_j, \ t=1,2,\dots,m,\label{eq:projection}
\end{equation}
where $\phi_j \in \mathbb{R}^{n^2}$ are some specific directions in graph space characterizing a suitable new base and $\boldsymbol\alpha(t) = [\alpha_1(t),\alpha_2(t),\dots,\alpha_r(t)]$ are the coordinates of the $t$-th snapshot in this lower-dimensional space, respectively. Observe that Eq.~\ref{eq:projection} trivially becomes an identity when $\phi_j$ are the canonical unit vectors in $\mathbb{R}^{n^2}$ and $r=n^2$, or when $\phi_j$ coincide with the eigenvectors of the covariance matrix ${\mathscr{S}}{\mathscr{S}}^\top \in \mathbb{R}^{n^2\times n^2}$. In this latter case, each $\phi_j$ is a suitable linear combinations of the canonical unit vectors. After de-flattening, $\{ \phi_j\}_{j=1}^{n^2}$ form a set of concrete adjacency matrices that we call network eigenmodes.
In the realm of fluid mechanics, this latter decomposition --traditionally applied to suitably flattened flowfield snapshots-- is called proper orthogonal decomposition (POD) \cite{lumley1967structure, taira2017modal}, and decomposes the fluid solution of some flow equation --e.g. Navier-Stokes-equations over a fixed domain-- in terms of a suitable combination of spatial coherent structures. 
Incidentally, POD shares well-known similarities with other equally popular concepts across the disciplines like the Karhunen–Loeve decomposition in signal processing or principal component analysis (PCA) in data science \cite{hastie2009elements}, and traces back to the theory of singular value decomposition \cite{horn2012matrix}. In the context of network theory, very recently Thongprayoon et. al \cite{thongprayoon2023embedding} proposed an analogous approach to use PCA for tie-decay temporal network embedding. As a matter of fact, the eigenvectors $\phi_j$ also provide the optimal\footnote{In the $L_2$ norm.} low-rank truncation of ${\mathscr{S}}$ \cite{horn2012matrix, eckart1936approximation}, i.e. for a given $r< n^2$, one cannot approximate better ${\bf a}(t)$ than with Eq.~\ref{eq:projection} where $\phi_j$ are the network eigenmodes. Furthermore, since covariance matrices are symmetric and semi-positive definite, these network eigenmodes $\{\phi_j\}$ can be ordered in terms of their associated eigenvalue $\{\lambda_j\}$ --which indicates the relative energy present in the mode--, and the spectral decomposition  of $\mathscr{S}\mathscr{S}^\top$ that reads $\mathscr{S}\mathscr{S}^\top = \sum_{i=1}^{n^2} \lambda_i \phi_i \phi_i^\top$ can be ordered such that $\lambda_1\geq\lambda_2 \geq \cdots \geq \lambda_{n^2}\geq 0$. Accordingly, the $r$ orthogonal directions in graph space coincide with the network eigenmodes with larger associated eigenvalue. Finally, the coordinates of each network snapshot in the new, $r$-dimensional space are found by projecting each network snapshot into the basis of modes
\begin{equation}
    \alpha_j(t)=\langle {\bf a}(t_j),\phi_j\rangle.
    \label{eq:coordinates}
\end{equation}
Incidentally, observe that when the number of nodes $n$ in each network snapshot is typically larger than the temporal resolution (more concretely, when $n^2 \gg m$), computationally speaking it is wiser to 
compute the
eigendecomposition of the covariance matrix of ${\mathscr{S}}^\top$, i.e. ${\mathscr{S}}^\top {\mathscr{S}} = \sum_{i=1}^m \lambda_i \psi_i\psi_i^\top$, 
because the covariance matrices ${\mathscr{S}}{\mathscr{S}}^\top \in \mathbb{R}^{n^2\times n^2}$ and ${\mathscr{S}}^\top {\mathscr{S}} \in \mathbb{R}^{m\times m}$ have the same set of non-null eigenvalues $\{\lambda_i\}_{i=1}^m$ (${\mathscr{S}}$ has at most rank $m$), and their associated eigenvectors $\{\phi_{i}\}_{i=1}^{n^2}$ and $\{\psi_i\}_{i=1}^m$ are related via
$$\phi_j = \frac{1}{\sqrt{\lambda_j}}{\mathscr{S}}\psi_j, \forall j \ \text{s.t.} \ \lambda_j\ne 0.$$
In a nutshell, the $m$ snapshots of a temporal network --which originally live in a space as large as the number of entries of the adjacency matrix-- can be optimally projected onto an $r$-dimensional space by leveraging on the eigendecomposition of a suitable temporal network's covariance matrix. The reconstruction error of the TN by using the compressed representation is controlled by the cumulated sum of the first $r$ eigenvalues.\\

Our second goal is to apply ideas from Koopman operator theory to the realm of temporal networks, much as dynamic mode decomposition (DMD) \cite{schmid2010dynamic} applies Koopman theory to fluid mechanical problems \cite{taira2017modal}. Given a --possibly nonlinear-- finite-dimensional dynamical system acting on a finite state space $\mathscr{X}$ (e.g. the latent graph dynamics which, adequately sampled, generates a temporal network formed by a sequence of weighted adjacency matrices), the Koopman operator is a {\it linear} operator acting on a (possibly infinite-dimensional) observable $g(\mathscr{X})$, such that $g(\mathscr{X}[t+1])=\mathscr{K}g(\mathscr{X}[t])$. {Koopman theory indeed shows that, for arbitrarily complex (nonlinear) dynamics, such linear operator always exist, i.e. any complex dynamics transform into a linear evolution in the space of the adequate observable $g(\mathscr{X})$. The drawback is that in general such adequate space is infinite-dimensional (a Hilbert space), and Koopman theoory does not prescribe a recipe to find the adequate observable $g(\mathscr{X})$ or a provably good finite-dimensional approximation\footnote{Observe that when the observable is identified with the state density --i.e. the marginal distribution of states--, the adjoint of the Koopman operator coincides with the transfer (Perron-Frobenius) operator. If instead of working in observable space, we directly work in state space, then a related procedure called Carleman linearisation can be exploited }. In complex systems it is therefore challenging, in general, to discover the coordinate transformations (observable) that linearize the dynamics \cite{brunton2016koopman}. Some general strategies are nonetheless available:} following the dynamic mode decomposition technique \cite{schmid2010dynamic, taira2017modal}, the simplest way to numerically approximate the (possibly infinite-dimensional, linear) Koopman operator $\mathscr{K}$ to the latent graph dynamics generating the temporal networks is probably to assume a linear evolution in the actual, finite state space ${\bf a}\in \mathbb{R}^{n^2}$, i.e. $g(\mathscr{X})={\bf a}$. In this case, the operator is the best-fit matrix $\mathscr{K}\in \mathbb{R}^{n^2 \times n^2}$ that advances the network trajectory, i.e. such that ${\bf a}(t+1)\approx  \mathscr{K}{\bf a}(t), \forall t=1,2,\dots,m$. From Eq.~\ref{eq:matrix}, we define two overlapping $\mathbb{R}^{n^2\times m-1}$ matrices 
\begin{eqnarray}
&&{\mathscr{S}_1}={\mathscr{S}}_{[1\dots m-1]}=[{\bf a}(1) \ {\bf a}(2) \ \dots \ {\bf a}(m-1)]; \nonumber \\ &&{\mathscr{S}_2}={\mathscr{S}}_{[2\dots m]}=[{\bf a}(2) \ {\bf a}(3) \ \dots \ {\bf a}(m)],
\label{eq:matrix}
\end{eqnarray}
such that least-squares fitting $\mathscr{K}$ is the (low-rank\footnote{Imposing low-rank is just one possible restriction. In the realm of quantum mechanics, we instead impose $\mathscr{K}$ to be a self-adjoint operator. In general, other physics-informed schemes can impose different structures. For instance, if energy is to be conserved, then $\mathscr{K}$ must be a unitary operator (reflection or rotation). In our case, if e.g. if networks are unweighted and the number of links is preserved along the dynamics, then $\mathscr{K}$ is a permutation matrix, i.e. we impose it to be an orthogonal binary matrix.}) solution of
\begin{equation}
    \mathscr{K} = \arg \min ||{\mathscr{S}_2} - {\mathscr{K}} {\mathscr{S}_1}||_F
    \label{eq:DMD_opti}
\end{equation}
 Since $\mathscr{K}$ is a linear operator, its eigenvectors correspond to coherent network structures (often called dynamic modes) and the eigenvalues describe their stylised growth, decay or oscillation that, all combined, characterise the linear approximation to the latent graph dynamics. Observe that if such latent dynamics is highly nonlinear, choosing 
$g(\mathscr{X})={\bf a}$ may yield a poor approximation to the true Koopman operator (e.g. time forecasting becomes a challenge). Accordingly, in that case one needs to look for a more sophisticated $g(\cdot)$ on which the linear action of $\mathscr{K}$ takes place. Invoking Takens embedding theorem\footnote{This theorem states that the attractor of high-dimensional nonlinear dynamics is diffeomorphic to the one reconstructed from any one-dimensional time series projection of the system, provided such projection is adequately embedded in sufficiently large time-delayed coordinates.} \cite{takens2006detecting, sauer1991embedology} one possible strategy to build such observable is to construct a time-delay embedding of ${\bf a}(t)$, i.e.  $g(\mathscr{X})=\tilde{{\bf a}}(t)=[{\bf a}(t), {\bf a}(t-1),\dots,{\bf a}(t-d)]$, where $d$ is the number of additional, time-delayed coordinates. Under this choice, the matrix-form of the trajectories $\tilde{\mathscr{S}}_X=[\tilde{{\bf a}}(1) \ \tilde{{\bf a}}(2) \ \dots \ \tilde{{\bf a}}(m-1)]$ is Hankel  \cite{brunton2017chaos, arbabi2017ergodic}. 
 While the $d=0$ case is intuitively simpler, note that the more general case $d>0$ is not really much more computationally harder, since while the dimensionality of $\mathscr{K}$ increases from $\mathbb{R}^{n^2\times n^2}$ to $\mathbb{R}^{n^2(1+d)\times n^2(1+d)}$, $\mathscr{K}$ is actually never explicitly computed nor diagonalised as we see below.

\medskip \noindent 
By construction, we have $\mathscr{K} \approx  \mathscr{S}_2 \mathscr{S}_1^{+}$, where $\mathscr{S}_1^{+}$ is the Moore-Penrose pseudoinverse of $\mathscr{S}_1$. The linear dynamics of the TN in observable space is thus fully described by the eigendecomposition of $\mathscr{K}$, 
 hence providing a data-driven way of assessing the intrinsic dynamics of temporal networks using the tools of linear dynamics. Now, instead of directly computing and then diagonalizing $\mathscr{K}$ --usually an ill-conditioned, and computationally hard problem for large networks, both for simple observables or time-delayed coordinate embedddings--, we leverage singular value decomposition (SVD) theory to compute the leading terms of its spectral decomposition as follows. First, ${\mathscr{S}}_1$ admits an SVD
${\mathscr{S}}_1= U \Sigma V^*,$ where, incidentally, the columns of $U$ are nothing but the network eigenmodes previously discussed in Eq.~\ref{eq:projection}. Similarly,
${\mathscr{S}}_2= \mathscr{K} U \Sigma V^*$. By further projecting into the network eigenmodes, we then have
\begin{equation}
  U^*{\mathscr{S}}_2 V \Sigma^{-1}= U^* \mathscr{K} U:= \tilde{\mathscr{K}},  
\end{equation}
where $\tilde{\mathscr{K}}\in \mathbb{R}^{m\times m}$ is in general much smaller than $\mathscr{K}$. $\tilde{\mathscr{K}}$ describes how the network eigenmodes evolve in time through the action of the linear dynamics.
We then diagonalize $\tilde{\mathscr{K}}=\sum_{i} \Lambda_i {\bf w}_i{\bf w}_i^\top$, where $W=[{\bf w}_1 \ {\bf w}_2 \ \dots]$ is eigenvector matrix and $\Lambda_i$ is the eigenvalue associated to ${\bf w}_i$. Now, by virtue of the properties of the SVD, the non-zero eigenvalues of $\mathscr{K}$ coincide with the eigenvalues of $\tilde{\mathscr{K}}$. The matrix containing the eigenvectors associated to the non-null eigenvalues of $\mathscr{K}$, which we denote $\Phi$, is finally calculated as \cite{schmid2010dynamic}
$$\Phi=\mathscr{S}_2 V \Sigma^{-1} W$$
(incidentally, when $\mathscr{S}$ admits a low-rank approximation (i.e. $r\ll \min(n^2, m)$), we use the reduced SVD instead of the full SVD). Finally, the growth/decay of each eigenvector or their frequency of oscillation corresponds, respectively, to the real and imaginary parts of $\log(\Lambda_i)$: eigenvalues inside the unit circle denote stable (exponentially decaying) modes, those outside the unit circle denote unstable (exponentially growing) modes, and eigenvalues which are roots of one yield pure oscillation with no growth or decay.

\begin{figure}[!htb]
\centering
\includegraphics[width=1.\columnwidth]{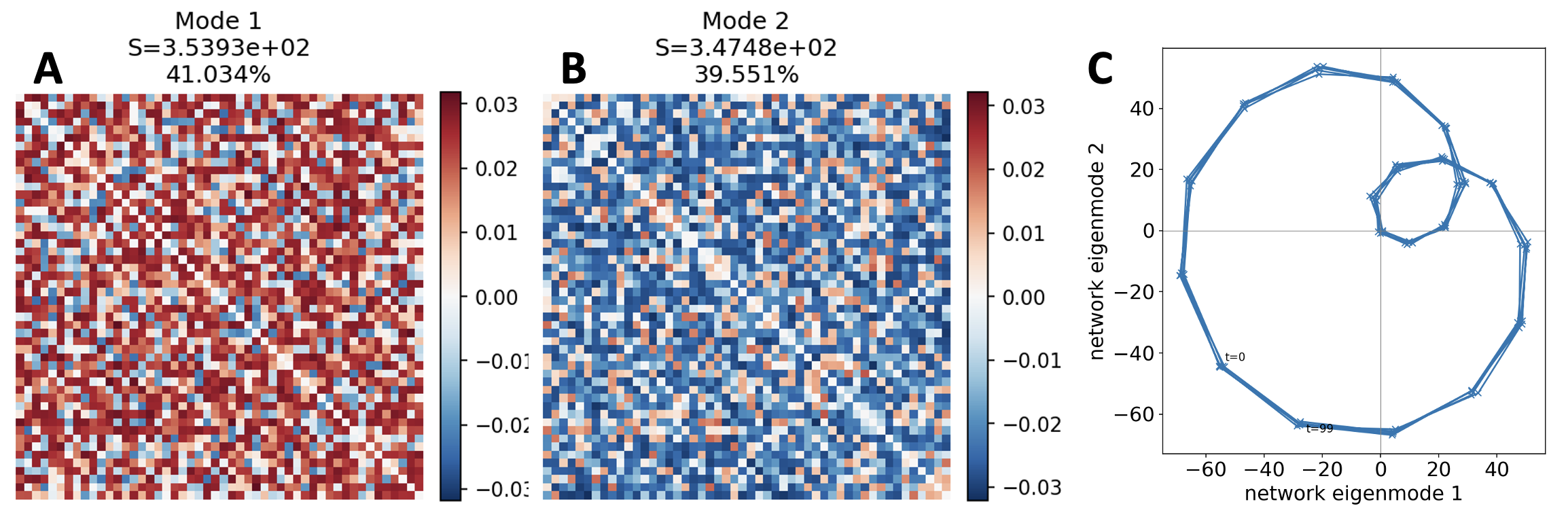}
\includegraphics[width=1.\columnwidth]{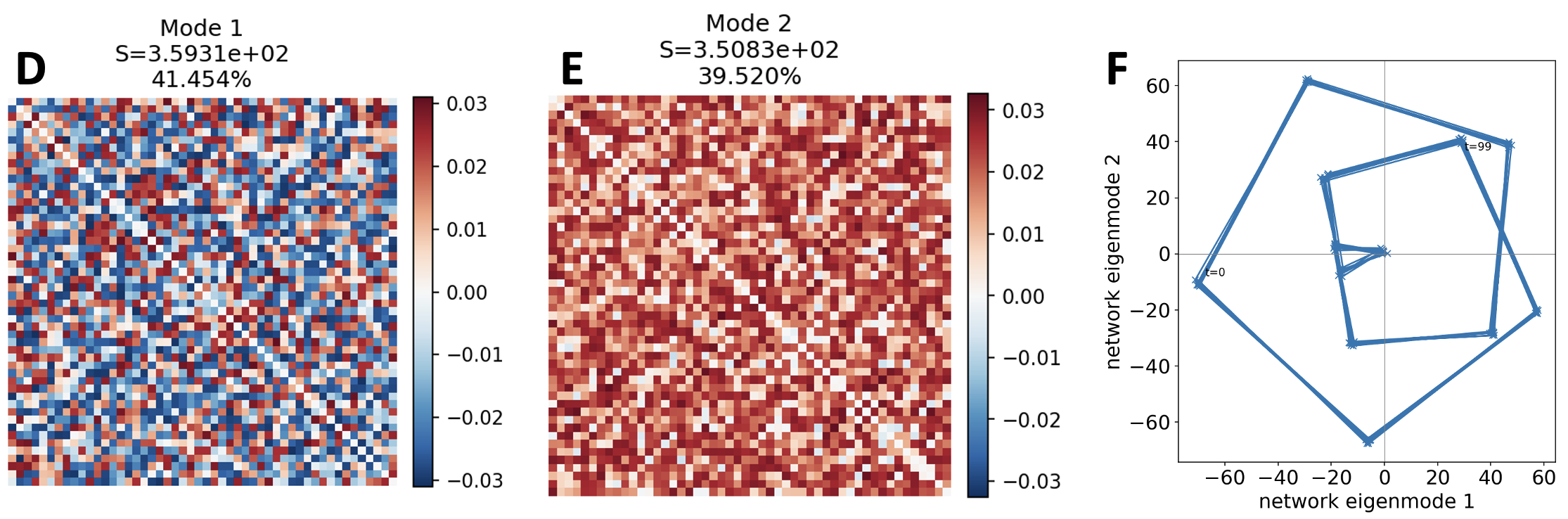}
\caption{{\small {\bf Compression of noisy, multiscale periodic TN dynamics.} (A,B,D,E) The first two network eigenmodes obtained from the eigendecomposition of the covariance matrix $\mathscr{S}\mathscr{S}^\top$ of a temporal network of $m=100$ network snapshots of $n=50$ nodes with asynchronous, noisy sinusoidal link dynamics of the form $A_{ij}(t)= \cos(2\pi t/T_1 + \eta_{ij}) + \cos(2\pi t/T_2 + \eta_{ij}) + \xi$,  with two concomitant periods $T_1$ and $T_2$ (see the text for details). The first two eigenmodes (out of a total of $n^2=2500$) already concentrate about $80\%$ of the variance. (C,F) Projection of the temporal network trajectory onto the space spanned by the first two network eigenmodes, resulting in a periodic orbit with period LCM($T_1,T_2$).}}
\label{fig:quasiperiod}
\end{figure}

\begin{figure*}[ht!]
\centering
\includegraphics[width=1.5\columnwidth]{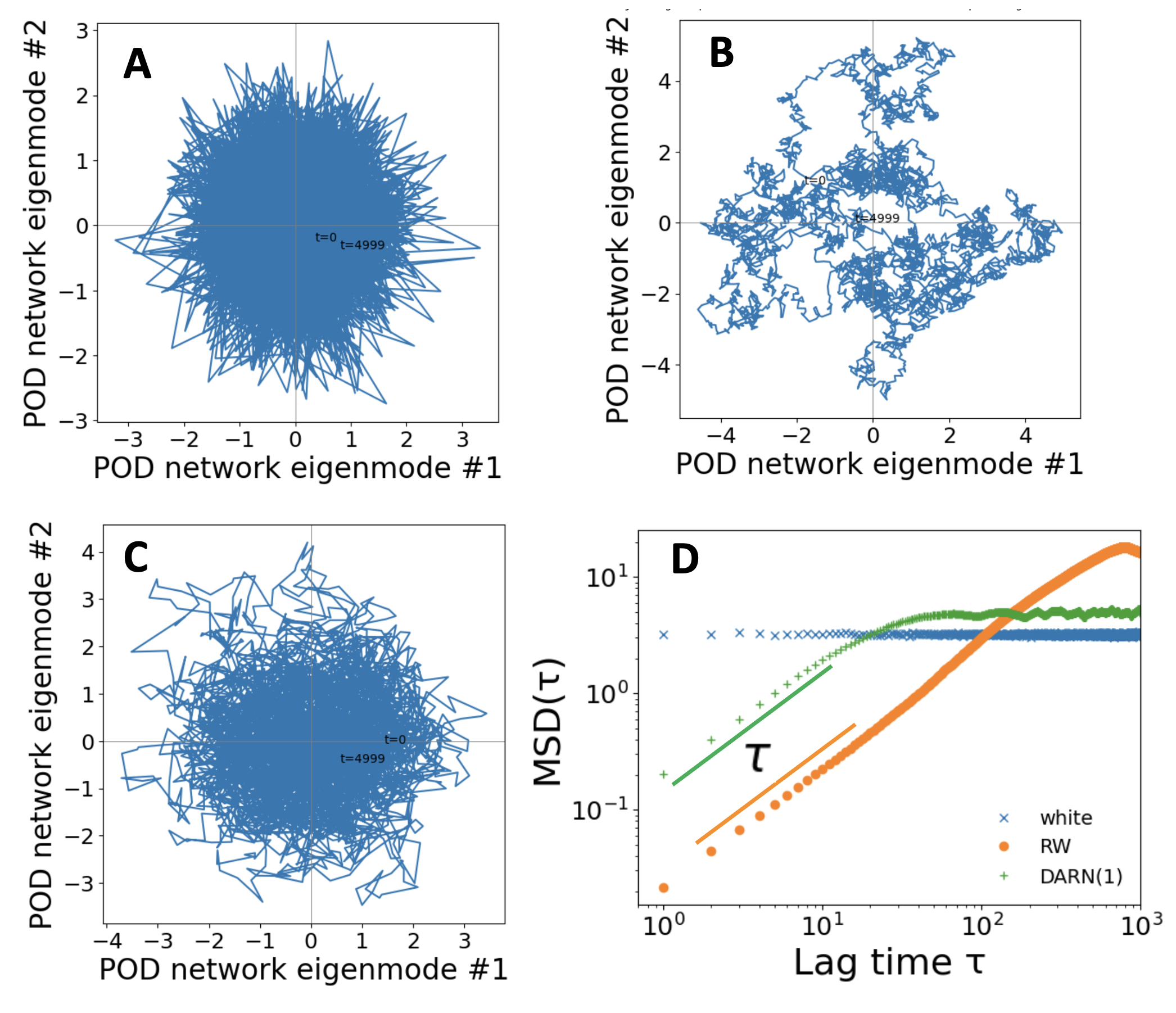}
\caption{{\small {{\bf Illustration of 2-dimensional POD projections of the dynamics of different TNs.} In every case TNs consist of $m=5000$ time snapshots, each snapshot having $n=20$ nodes. Projections are made into the two leading POD network eigenmodes (out of the total of 400 eigenmodes). Dynamics include: (A) white temporal network, (B) A Random Walk TN, (C) a DARN(1) process. The energy stored in the first two eigenmodes amounts to (A) $1.5\%$, (B) $10\%$, (C) $2.6\%$. (D) Mean square displacement of the projected trajectory, for the three projections. Results match the behavior of MSD for iid ($\text{MSD}(\tau)\sim \tau^0$), random walks ($\text{MSD}(\tau)\sim \tau$) and autorregressive processes ($\text{MSD}(\tau)\sim \tau$ at short times, and $\text{MSD}(\tau)\sim \tau^0$ at long times), hence suggesting that the projections preserve the dynamical fingerprint of the original TN dynamics, even if the eigenmodes accumulate very modest variance.}}}
\label{fig:eigens}
\end{figure*}

\section*{Results} \noindent {\bf Compression via POD --} We have constructed generative temporal networks with a suite of different intrinsic dynamics, including (i) white noise TNs, (ii) different types of noisy periodic dynamics, (iii) autorregressive network processes with memory, (iv) network chaotic dynamics \cite{lacasa2025scalar,lacasa2022correlations,annalisa} and (v) an empirical temporal network, see {Methods for details}. Results overall indicate that the intrinsic dynamics of the TN is systematically well preserved in the orbits projected into the leading POD network eigenmodes, yielding a substantial TN compression for all models except for white noise TN, which by construction cannot be compressed. The coordinates of these low-dimensional orbits are computed from Eq.~\ref{eq:coordinates}. Reconstruction of the high-dimensional TN dynamics from the compressed orbits is obtained by using Eq.~\ref{eq:projection}.\\ 

\medskip \noindent 
{Let us now unfold a detailed description of all the analysis step by step.} As an initial check, we consider a TN built by concatenating several times the same (iid) sequence of $T$ Erdos-Renyi network snapshots, so that the resulting TN has period $T$ and is fully uncorrelated at lags different than $T$ or its multiples \cite{lacasa2022correlations}. After computing the eigendecomposition of $\mathscr{S}\mathscr{S}^\top$, we found that the first $T-1 \ll n^2$ network eigenmodes concentrate about $100\%$ of the variance. Furthermore, the trajectory of the TN projected in the first two network eigenmodes is itself a periodic orbit of $T$ points, yielding a massive compression with small reconstruction error.\\
Subsequently, we now build a more challenging periodic TN model, where
the adjacency matrices have asynchronous and multiple-period noisy link dynamics $A_{ij}(t)= \cos(2\pi t/T_1 + \eta_{ij}) + \cos(2\pi t/T_2 + \eta_{ij}) + \xi$, where $T_1$ and $T_2$ are the two intrinsic periods, $\eta_{ij}\sim \textsc{uniform}(0,1)$ is a quenched noise that desynchronises the dynamics of each link, and $\xi\sim \textsc{N}(0,\sigma)$ is a Gaussian dynamical noise. Fig.~\ref{fig:quasiperiod} displays the first two POD network eigenmodes (panels A and B, D and E) and the projection of the original TN into the space spanned by these eigenmodes (panels C and F) for a TN of $m=100$ time snapshots and $n=50$ nodes per snapshot following this dynamics with $\sigma=0.5$ and $T_1=20, T_2=10$ (panel C) and $T_1=6, T_2=4$ (panel F). The periodic nature of the TN is again well captured by the orbit in the 2-dimensional network eigenmode space, which is itself quasi-periodic (periodic for noise $\sigma\to0$) with a quasi-period equal to the lowest common multiple (LCM) of $T_1$ and $T_2$ ($\text{LCM}(20,10)=20, \text{LCM}(6,4)=12$). 

\medskip \noindent
{As a complementary illustration of two-dimensional projections, in Fig.~\ref{fig:eigens} we plot the two-dimensional orbits (projected into the two leading POD network eigenmodes) associated to temporal networks with $n=20$ nodes, $m=5000$ snapshots and generated by {different three types of stochastic dynamics (see methods for details): (A) a white (iid) sequence of network snapshots, (B) an unbiased random walk in graph space, and (C) a discrete autorregressive network process of order 1 DARN(1), which is the discrete, temporal network analog of an Ornstein-Uhlenbeck process \cite{williams2022shape}}. Visually, the projections are on qualitative agreement with what we would expectat for the iid (no structure) and the random walk case, although note that the autorregressive case is, visually, unexpectedly similar to the white noise case, despite the fact that autorregressive models have short-term memory. To quantitatively check whether the dynamical properties of the TNs are preserved in the projection,
panel (D) plots in log-log the mean square displacement scaling $\text{MSD}(\tau)$ as a function of time $\tau$, computed on the 2-dimensional projections for all three TN dynamics considered in this figure.  It is well known that $\text{MSD}(\tau)$ is constant for 2-dimensional white noise, scales linearly with time for unbiased random walks and shows linear scaling for short times while saturates to a constant value for long times in the case of autorregressive processes. Panel (D) precisely finds this behavior, validating that the projection preserves the properties of the original dynamics, even if the variance contained by the first two POD modes is quite modest.} 



\begin{figure}[htb!]
\centering
\includegraphics[width=1.
\columnwidth]{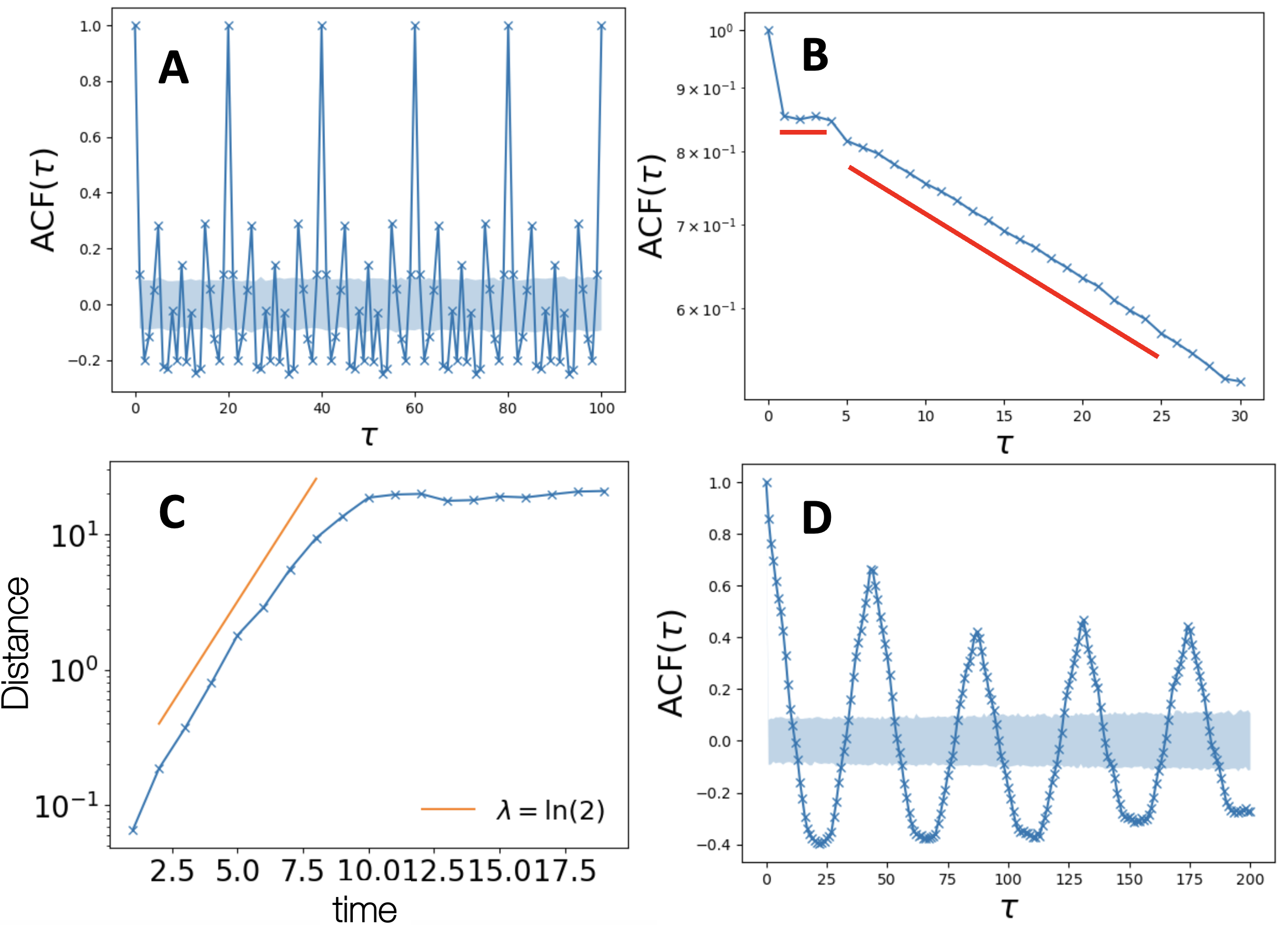}
\caption{{\small {{\bf Characterisation of TN dynamics from their scalar POD-projection.} (A) ACF($\tau$) of the projected network trajectory into the leading eigenmode ($6\%$ variance) for a periodic TN with period $T=50$ of $m=500$ time snapshots, where each snapshot has a fixed number of nodes $n=50$ and links (i.e. its periodicity is not related to a periodically fluctuating link density). (B) Same as (A), but applied to a DARN(4) model for a TN of $m=10^4$ time snapshots and $n=10$ nodes per snapshot (the leading eigenmode only captures $0.8\%$ of the variance). The panel shows the autocorrelation function $\text{ACF}(\tau)$ in semi-log, correctly retrieving the stylised shape of an autorregressive process (flat autocorrelation up to memory order, followed by an exponential decay), even if the leading eigenmode only captures $0.8\%$ of the variance. (C) Semi-log plot of the ensemble-averaged distance of nearby network initial conditions in the projection of the chaotic TN network (first POD eigenspace, $67\%$ variance) over time, computed via the Wolf method (see methods), applied to a chaotic temporal network of $n=50$ nodes and $m=1000$ time snapshots generated by the dictionary trick from the fully chaotic logistic map (see methods). The distance increases exponentially with a slope in good agreement with the (network) Lyapunov exponent $\lambda=\ln 2$. (D) Empirical temporal network of proximity in a workspace, over roughly 4 days (discarding transients, 163 snapshots where each snapshot time-aggregates links over a window of $1987.08$). In order to remove any trace of link density fluctuations (which could flag easy-to-spot periodicity), the network has been dynamically polluted with substantial amounts of noise by adding many links at random to each snapshot, so as to keep each snapshot with the same number of links. Despite such contamination, the autocorrelation function of the network trajectory in the leading POD mode projection (only $0.9\%$ variance) still captures a clear periodic backbone with period $T=44$, corresponding to 24.3 hours, i.e. daily periodicity.}}}
\label{fig:chaotic}
\end{figure}

\medskip \noindent {By construction, observe that $n^2$ eigenmodes are needed in the worst case scenario to fully reconstruct the TN trajectory, but often the key dynamical fingerprints can be retrieved by projecting the trajectory in a few network modes, as is suggested by the results in Fig~\ref{fig:eigens}. In order to complement these analysis, we now explore how projection into the {\it leading} network eigenmode generates signals (scalar time series) that already preserve a large deal of the high-dimensional dynamics. 
Results are presented in the four panels of Fig.~\ref{fig:chaotic}.
In this figure, we have further considered four scenarios where we have access to a `ground truth': (A) a periodic TN built by concatenating the same sequence of $T$ iid Erdos-Renyi graphs over and over, where the number of links is kept fixed (i.e. period $T$ is not attributed to link density fluctations), (B) an autorregressive TN dynamics of relatively high order (DARN(4)), (C) a chaotic TN dynamics generated by the dictionary trick \cite{lacasa2022correlations, annalisa} with known network maximum Lyapunov exponent $\lambda = \ln(2)$ (see methods), and (D) an empirical social temporal network from proximity contacts in a workplace environment (see methods). These TN dynamics are projected into the leading POD network eigenmode and their characterisation is made in terms of the temporal autocorrelation structure ACF($\tau$) of the projected signal (panels A,B and D) and via a derivation of the maximum Lyapunov exponent from (panel C). Panel (A) clearly detects periodicity at the correct $\tau=T$, despite the leading POD network eigenmode accounting for a not impressive $6\%$ of the variance. 
Panel (B), which depicts the signal's ACF($\tau$) of the signal projected from the DARN(4) TN (in this case the leading network eigenmode only captures $0.8\%$ of the variance) recovers the stylised form of the ACF for autorregressive processes: a flat ACF($\tau$) for $\tau\le p$ (where $p$ is the order of the memory, here $p=4$), followed by an exponential decay\footnote{Note that in this case the number of necessary snapshots is much larger, this was needed to obtain a clear exponential decay}. In Panel (C) we consider a TN of $m=1000$ snapshots with $n=50$ nodes per snapshot which follows a chaotic evolution with theoretical network maximum Lyapunov exponent $\lambda=\ln 2$, generated by the dictionary trick (methods). The projection into the leading POD network eigenmode preserves a considerable amount of variance (67\%) of the original TN, as the actual chaoticity of this network is of low-dimension (effectively one dimensional). To validate in this case that the scalar projection captures the chaotic fingerprint, we check whether initial close conditions deviate exponentially fast under the action of the dynamics directly in the time series via Wolf's method (see methods). Accordingly, Panel (C) plots, in semi-log scales, the expansion of nearby initial conditions over time (averaged over 100 initial network conditions picked at random from the TN). The plot clearly shows an exponential separation with a slope that is on agreement with $\lambda=\ln 2$. Finally, panel (D) plots again the ACF($\tau$) applied now to an empirical proximity social temporal network of co-workers in a workplace, which by construction has an intrinsic daily periodicity. To make the detection more challenging, this TN has been contaminated with very larges amounts of dynamical noise --adding links at random in each snapshot-- so as to keep the link density constant for every snapshot, thereby removing any periodic effect related to a possible periodic link density (the original temporal network had an average of 31 links per snapshot, whereas the polluted temporal network has 515 links per snapshot).
The processed TN is projected into its leading network eigenmode ($0.9\%$ of the variance), and the ACF clearly shows a periodic structure, with a period $\tau = 44$. Observe that in this TN each snapshot aggregates the link activity over a time window of about 1987 seconds, so the period is therefore $44\times 1987/3600 \approx 24.3$ hours, close to the daily periodicity expected in such a social network.}

\begin{figure*}[htb!]
\centering
\includegraphics[width=1.7\columnwidth]{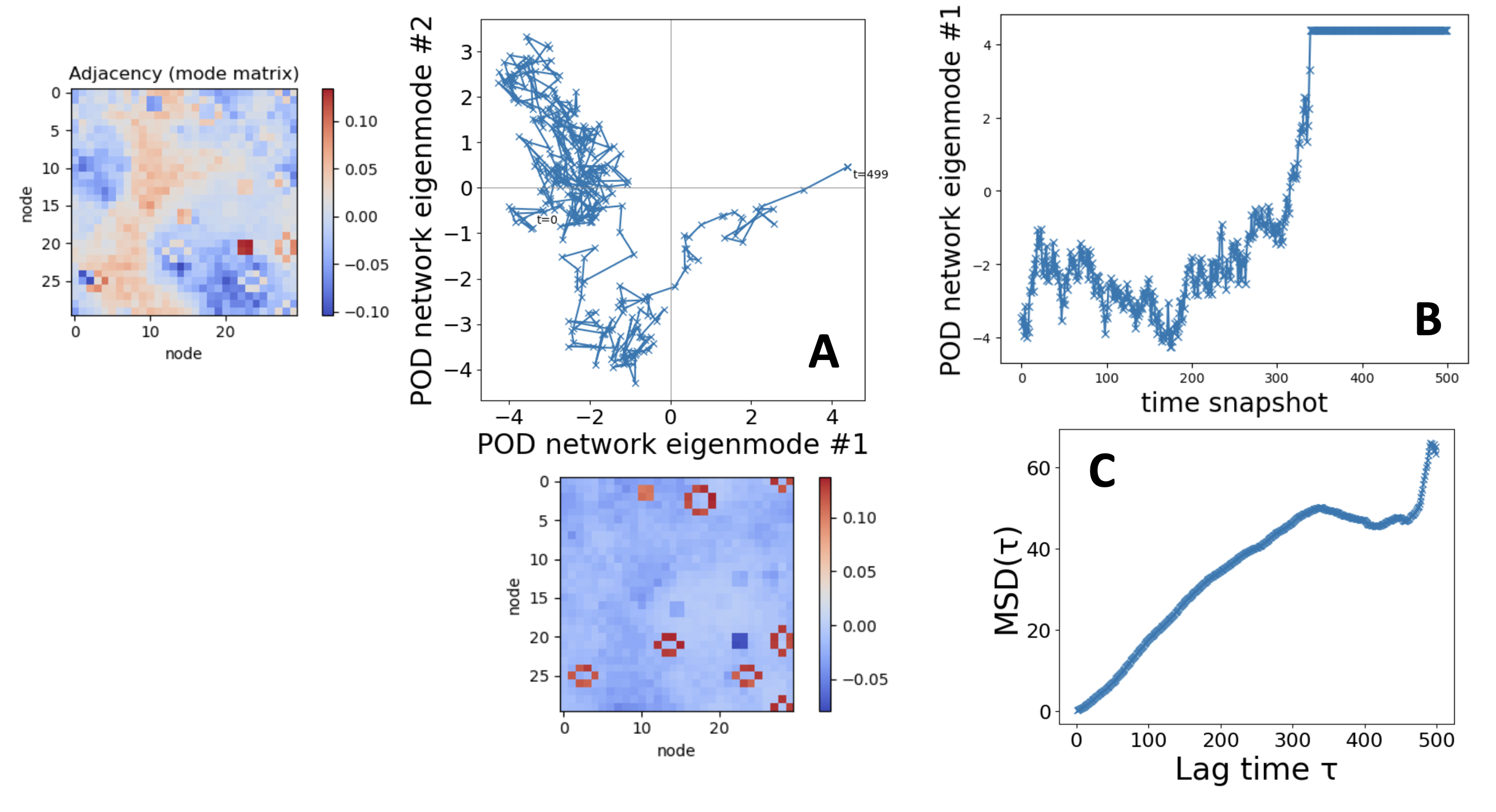}
\caption{{\small {{\bf Conway's Game of Life as a temporal network. } We set $n=30$ nodes (akin to a $30\times 30$ lattice) and run Life for $m=1000$ time steps (network snapshots). (A) Evolution of Life in the space spanned by the first two network eigenmodes (the actual eigenmodes are depicted at the sides, and account for $15\%$ and $4\%$ variance, respectively). This trajectory is similar to a random walk, see Fig.~\ref{fig:eigens}B. The mean-square displacement is plotted in Panel (C), showing a linear scaling regime up to a cut-off time. (B) Projecting into the leading eigenmode detects the onset of the fixed point in the dynamics at $t\approx 350$, and this coincides with the cut-off of the linear MSD scaling.}}}
\label{fig:conway}
\end{figure*}

\medskip
\noindent As a final illustration, we consider an exotic `experimental' application that aims to showcase the generality of the approach: we identify the successive temporal states of a 2D cellular automata (2DCA) as the adjacency matrices (snapshots) of a temporal network. Concretely, we implement Conway's Game of Life \cite{Izhikevich:2015} (Life for short) --a Turing-complete 2DCA \cite{rendell2002turing}--. {Life follows a deterministic dynamics in a finite state space, and thus cannot strictly evidence chaotic behavior, as it eventually reaches a periodic attractor made of still patterns (blocks, beehives, boats, etc), and oscillators (blinkers, toads, pulsars, etc). However, transient dynamics towards the attractor can be long and seemingly unpredictable, and include the appearance of coherent structures like spaceships or gliders. For illustration, we run a single realization of Life on of $30\times 30$ (i.e., $n=30$ nodes) and $m=500$ evolution steps. The first two network eigenmodes and the projection of Life --interpreted now as a TN-- is displayed in Fig.~\ref{fig:conway}A. The low-dimensional trajectory is similar to the one found for a stochastic TN (linear scaling of the MSD, Panel C), as expected since Life transients are unpredictable until reaching one of its attractors. Convergence to this attractor is clear from Panel B which plots the projection into the leading network eigenmode, observe in particular that the time the dynamics reach a fixed coincide with the time when the MSD linear scaling of the 2d projection stops.}

\begin{figure}[htb!]
\centering
\includegraphics[width=1.\columnwidth]{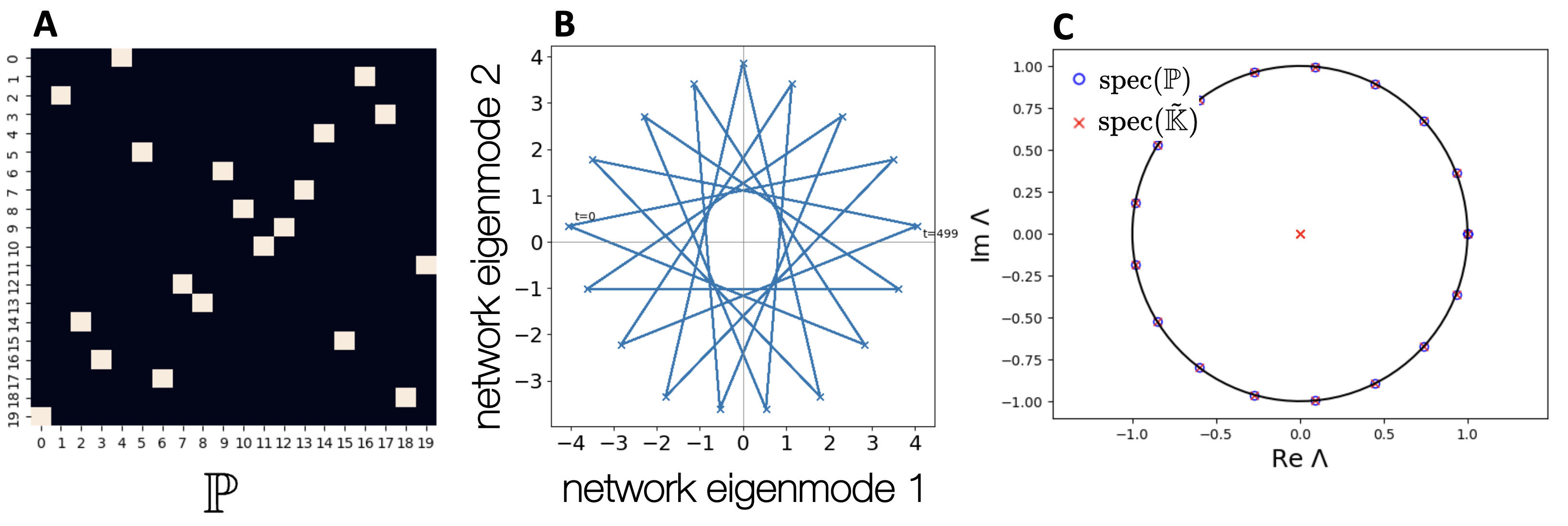}
\caption{{\small Permutation TN dynamics ${\bf A}(t+1)=\mathbb{P} {\bf A}(t)\mathbb{P}^\top$ for a network of $n=20$ nodes over a total of 500 time snapshots. $\mathbb{P}$ is a randomly chosen permutation matrix which induces a cyclic dynamics of period 17.  
(A) $20\times 20$ permutation matrix $\mathbb{P}$. (B) Projection of the TN dynamics onto the plane spanned by the first two network eigenmodes. The trajectory is periodic, mimicking the periodic cycle induced $\mathbb{P}$, recovering the periodicity. Observe that such periodicity can already be retrieved from the periodic behavior of the projection in the first POD network eigenmode (which accumulates about 9\% of the variance).
(C) Spectrum of $\tilde{\mathscr{K}}$ and $\mathbb{P}$. The leading $n$ eigenvalues of $\tilde{\mathscr{K}}$ coincide with those of $\mathbb{P}$, being roots of 1 that yield no growth or decay of dynamic modes, as expected for a permutation operator.}}
\label{fig:permut}
\end{figure}
\begin{figure}[htb!]
\centering
\includegraphics[width=1.\columnwidth]{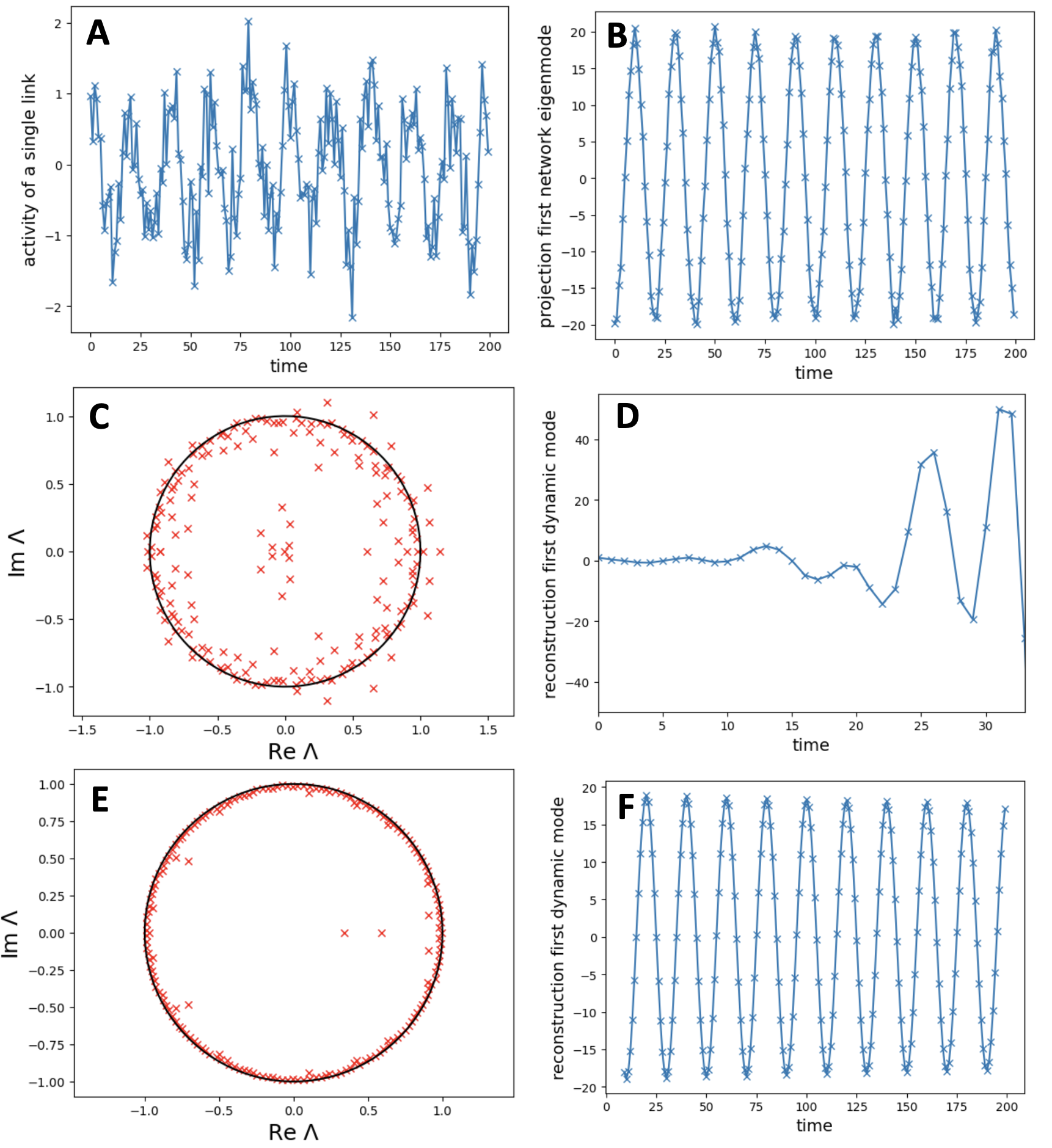}
\caption{{\small {Full analysis on a noisy period temporal network of $m=200$ snapshots with $n=20$ nodes and link dynamics $A_{ij}(t)=\cos(2\pi t/T) + \xi$, where $\xi \sim N(0,0.5)$ and $T=20$. (A) $A_{ij}(t)$ for a concrete link, showing a noisy periodic evolution with period $T=20$. (B) Projection of the whole TN onto the first (POD) network eigenmode, that accumulates over 67\% variance, recovering the periodic backbone. (C) Spectrum of $\tilde{\mathscr{K}}$ obtained from a standard DMD approach. The norm of ${\bf A}(t)$ varies over time, what leads to the emergence of spurious unstable modes in $\tilde{\mathscr{K}}$ (eigenvalues outside the unit circle). (D) TN reconstruction is affected by unstable dynamic modes. (E) Like (A), but after delay-embedding each snapshot $\tilde{\bf a}(t) = [a(t),\dots, a(t-10))]$, thereby building a more sophisticated observable $g(\mathscr{X})$ where the linear dynamics is assumed to take place. In this case the spurious unstable modes disappear completely, and reconstruction is now possible (F).}}}
\label{fig:Hankel}
\end{figure}

\medskip 
\noindent {\bf Network stability via DMD --}
Moving on, we now illustrate the approximation of Koopman's operator $\mathscr{K}$. To that aim, we initially generate a TN consisting in iterating a permutation of indices in the adjacency matrix ${\bf A}(t+1)=\mathbb{P}{\bf A}(t)\mathbb{P}^\top, t=1\dots m, m=50$ where $\mathbb{P}$ is a particular $n\times n$ permutation matrix with $n=20$ chosen at random, see Fig.~\ref{fig:permut}A for an illustration. By definition, such dynamics preserve the number of links. {Since this is a conjugation, it is a special case of a linear operator.}
The TN trajectory, as projected in the two leading POD network eigenmodes, is shown in  Fig.~\ref{fig:permut}B, finding a periodic orbit whose period coincides with the specific permutation cycle generated by iteration of $\mathbb{P}$. Then, in Fig.~\ref{fig:permut}C we compare the spectrum of $\mathbb{P}$ with the spectrum of $\tilde{\mathscr{K}}$. All eigenvalues lie in the unit circle, i.e. there are no growing or decaying dynamic modes --as expected sin permutation induces periodic (cyclic) dynamics with fixed number of links (conservative)-- and coincide
our approximation to the Koopman operator is in this case exact, as the actual latent graph dynamics is already linear in the natural observable. {A similar phenomenology is found for a generic linear dynamics of the form ${\bf A}(t+1)=\mathbb{P}{\bf A}(t)$: if $\mathbb{P}$ is a permutation matrix, even if the dynamics is not akin to node relabelling (as the action only permutes rows), it generates periodic oscillations with the same period as the permutation cycle induced by $\mathbb{P}$, whereas for a generic matrix $\mathbb{P}$, its spectral properties determine the stability of the dynamics and this is exactly recovered from $\tilde{\mathscr{K}}$.}

\medskip \noindent
As a final analysis, we consider now a simple sinusoidal TN dynamics where each individual link evolves following $A_{ij}(t)=\cos(2\pi t/T)+\xi$, with intrinsic period $T=20$ and $\xi \sim N(0,0.5)$ is an extrinsic Gaussian white noise. The network is undirected and thus we set $A_{ij}=A_{ji}$. This generates a noisy periodic TN, but the norm of each adjacency matrix severely fluctuates over time. {For illustration, Fig.~\ref{fig:Hankel}A shows the time evolution of the weight of an individual link, showing a noisy periodic dynamics with a periodicity of $T$. Fig.~\ref{fig:Hankel}B depicts the projection of the TN onto the first network eigenmode given by the POD, which accumulates about 67\% of the TN variance. The periodicity is clear and the noise has been filtered. Then we consider the approximation of the Koopman operator from the standard DMD. Fig.~\ref{fig:Hankel}C shows the spectrum of $\tilde{\mathscr{K}}$, where we see that a few unstable modes emerge (eigenvalues outside the unit circle). The presence of these (spurious) modes ruin down the reconstruction, see Fig.~\ref{fig:Hankel}D where reconstruction from the leading (unstable) dynamic mode blows up. Interestingly, these (spurious) unstable modes disappear completely if we extend the DMD approach by conducting an appropriate Takens embedding approach (so as to build a more sophisticated observable $g(\mathscr{X})$), see Fig.~\ref{fig:Hankel}E-F where we used a delay dimension $d=10$. In this latter case, the spurious unstable modes have disappeared, and the periodic dynamics can be reconstructed, indicating that the stability of the original TN dynamics needs to be assessed in this more sophisticated observable.}

\section*{Discussion} {This works explores the potential application of data-driven methods in fluid dynamics to the field of network science, particularly temporal networks. } We have proposed two different eigendecompositions applicable to temporal networks, each of them yielding useful information that can be used for a purely data-driven compression (low-dimensional representation) and spectral characterization of the latent graph dynamics, respectively. The former, being a simple extension of POD to the realm of temporal networks, is aligned with recent strategies for temporal network embeddings \cite{thongprayoon2023embedding, lacasa2025scalar}, and our analysis indicates that even strong lossy compression of the TN preserves key dynamical fingerprints of the original dynamics.
Further work should explore its applicability across areas including social and biological systems, where (temporal) networks are pervasive.\\
With regard to approximating the Koopman operator, the method provides useful spectral characterization of a linearization of the latent graph dynamics when such dynamics is close to linear and works well when the number of links is approximately constant over time. {This was for instance illustrated in the case of a temporal network generated by successively permuting an original adjacency matrix, where the dynamics is linear and the Koopman operator is therefore easily found, thereby allowing for an exact reconstruction of the (high dimensional) temporal network dynamics from the DMD modes.} Now, for TNs where the number of links (e.g. the adjacency matrix norm) severely fluctuates over time, the basic DMD implementation  captures (spurious) unstable dynamic modes that quickly govern the dynamics and ruins long-term forecasting. In this case, a more accurate dynamical reconstruction is found by delay-embedding the network trajectory so as to build a higher-dimensional observable.\\
On a technical note, three comments are in order. First, observe that graphs are invariant under node relabelling, whereas adjacency matrices aren't. This can be seen as a limitation, as any a priori obscure node labelling can hinder the {visual interpretability of the patterns observed in the different network eigenmodes, although the quantitative information one can retrieve from either decomposition --for compression, stability analysis, or reconstruction-- is in itself not affected, as reported in the paper. Second, both methods rely on nodes maintaining their labels accross snapshots, and thus they are currently only designed for labelled temporal networks. Adequate extensions of these methods for unlabelled networks are thus needed \cite{caligiuri2025characterizing}. 
Third, observe also that while POD does not directly use temporal ordering information of network snapshots for the TN decomposition --similar to other TN embedding approaches \cite{thongprayoon2023embedding, lacasa2025scalar}--, eigendecompositions based on DMD and its extensions assess the dynamical stability in observable space and thus these do consider the temporal ordering.}

\medskip {To conclude, this paper proposes a potentially fruitful application of fluid-dynamic techniques to temporal network analysis. Our proof of concept explores how TNs can be compressed --i.e. its dynamic evolution be projected in a low-dimension eigenspace-- and how their dynamical stability and associated reconstruction can be studied with classical fluid-dynamic approaches such as POD or DMD. Future work should explore how refinements of basic POD methods --e.g. kernel POD or robust POD-- can help improve the compression problem while accounting for highly nonlinear or noisy temporal networks, while further refinements of the basic DMD or its extended version discussed here as well --e.g. kernel DMD, Total Least Squares DMD, etc-- can help with the stability/forecasting problem for highly nonlinear, non-conservative or noisy TNs. The comparison of these suite of fluid-dynamic methods against more traditional techniques used in the analysis of TNs is also needed to assess their practical interest. Notably, observe that true Koopman mode instabilities emerging in the data-driven linearization of the latent graph dynamics can also benefit from control-theoretic techniques, which in turn provide concrete interventions in the temporal network aimed to preserve its stability. We also hope this paper will stimulate further research that addresses, from a fluid-dynamic angle, other properties of TNs besides compression and stability/forecasting. Finally, and further building on the cross-pollination rationale, it might be interesting to extend such methods not only to temporal networks, but to complex systems that are naturally described as sequences of 2D field snapshots. This was already considered here e.g. in the application to 2D cellular automata, but further examples include lattice-based spin-like systems (Ising model, voter models, sandpile models of self-organised criticality) or lattice-based pattern formation dynamics}.\\

\noindent {\bf Methods --} {Here we provide some details on some of the generative models for temporal networks which have been used in this work.}

\medskip \noindent 
{{\it (i) iid TN}: a sequence of Erdos-Renyi graphs sampled at random forms a `white' temporal network, in the sense of being a time series of graphs with Delta-distributed network autocorrelation function \cite{lacasa2022correlations}.}

\medskip \noindent 
{{\it (ii) RW TN}: starting from an Erdos-Renyi graph, we build an ordered sequence of graphs where the next graph in the sequence is built by choosing a site $(i,j)$ in the matrix and flipping its entry $A_{ij} \to 1 - A_{ij}$.}

\medskip \noindent 
{{\it (iii) DARN(p) TN}: in the Discrete Autorregressive Network model DARN(p) \cite{williams2019effects, williams2022shape}, the dynamics of each link $\ell_t$ in the $t$-th graph snapshot is constructed independently, such that with probability $q$, $\ell_{t+1}$ samples uniformly from its past $p$ states, and with probability $(1-q)$, it assigns a Bernoulli trial with probability $y$. 
In other words, when the link update is random, we flip a biased coin and assign the entry $1$ (link present) with probability $y$ and the entry $0$ (link absent) with probability $1-y$). Such process generates a non-Markovian network trajectory with memory order {$p$} and a characteristic network autocorrelation function \cite{lacasa2022correlations}.
In this work we fix the parameters $q=0.9, \ y=0.5$.}

\medskip \noindent 
{{\it (iv) low-dimensional chaotic TN via dictionary trick}: We initially consider a one-dimensional time series, for instance the one 
generated by the fully chaotic logistic map $x_{t+1}=4x_t(1-x_t)$. This is an interval map $x\in [0,1]$, so the algorithm proceeds by generating a time series $(x_t)_{t=1}^T$, and symbolising the signal after homogeneously partitioning the interval $[0,1]$ into $Q$ equally-sized cells. In parallel, we construct a dictionary of $Q$ networks $\{G[q]\}_{q=1}^Q$ (observe that this is just a set of networks, not a temporal network). This `network dictionary' is built sequentially: starting from an initial (e.g. Erd{\H{o}}s-R\'enyi) network of $n$ nodes $G[1]$, in each step of the process a unique link rewiring is performed. Iterating such process builds $G[2]$, $G[3]$, etc. Now, such rewiring needs to follow two strict rules: (i) one cannot select a link which had already been inserted from a previous rewiring, and (ii) the new link cannot be inserted in a place which previously had a link that had eventually been rewired. By following these two rules, {one can prove that the sequence of generated networks in the dictionary is metrical}: any two $G[s]$ and $G[t]$ are precisely $t-s$ rewirings apart, so ${\left\| G[s] - G[t] \right\|}\propto|t-s|$ \cite{lacasa2022correlations, annalisa}. Once the network dictionary is built, each cell of the interval $[0,1]$ is matched with a network of this dictionary, so that the first cell is assigned $G[1]$, the second cell is assigned $G[2]$, and so on. Finally, each point of the time series $x_t$ can now be symbolised as a network which we label $G_t$. The temporal network $(G_t)_{t=1}^T$ constructed in this way inherits the dynamical properties of $(x_t)_{t=1}^T$, in particular the same Lyapunov exponent $\lambda=\ln 2$.}

\medskip \noindent {{\it Wolf method \cite{wolf1985determining}}: To estimate sensitive dependence on initial conditions directly from a time series $x(t)$, Wolf's method proceeds to find recurrences in such time series, i.e. pairs of points $x(k), x(k')$ which are far apart in time (i.e. $k'\gg k$) but very close in space $|x(k) - x(k')|\ll 1$. These pairs are considered as nearby initial conditions, and the algorithm tracks the successive positions over time to build a distance $d(\tau)=|x(k+\tau) - x(k'+\tau)|$. A system showing sensitive dependence on initial conditions will show $d(\tau)\propto \exp(\lambda \tau)$. In order to make the calculations independent on the initial condition, one can average results over such an ensemble of different initial conditions (different pairs).}

\medskip \noindent {{\it (v) Empirical social temporal network}: we consider a 
social temporal network \cite{GÉNOIS_VESTERGAARD_FOURNET_PANISSON_BONMARIN_BARRAT_2015, genois2018can} from the SocioPatterns collaboration
(\url{sociopatterns.org}), constructed from proximity contacts between co-workers in a workplace environment, over roughly 4 days (discarding transients, 163 snapshots where each snapshot time-aggregates links over a window of $1987.08$). In order to remove any trace of link density fluctuations (which could flag easy-to-spot periodicity), the network has been dynamically polluted with substantial amounts of noise by adding many links at random to each snapshot, so as to keep each snapshot with the same number of links (original temporal network had an average of 31 links per snapshot, polluted temporal network has a fixed 515 links per snapshot).}


\medskip
\noindent {\bf Data and code availability --}
The datasets and code generated during the current study are available from the corresponding author on reasonable request and will be published in \url{github.com/lucaslacasa} upon publication.

\medskip
\noindent {\bf Author contributions --} L.L. conceived research, performed research and wrote the paper. 

\medskip
\noindent {\bf Funding --} The author acknowledges partial support from project CSxAI (PID2024-157526NB-I00) funded by MICIU/AEI/10.13039/501100011033/FEDER, UE, project Maria de Maeztu CEX2021-001164-M funded by the MICIU/AEI/10.13039/501100011033, and from the European Commission Chips Joint Undertaking project No. 101194363 (NEHIL).\\


\end{document}